\documentclass[article,9pt]{IEEEtran}
\usepackage{amsmath}
\usepackage{amsfonts}
\usepackage{amssymb}
\usepackage{setspace}
\usepackage[latin1]{inputenc}
\usepackage{graphicx}
\usepackage{amsthm}
\usepackage{color}
\usepackage{cite}
\usepackage{bm}
\usepackage{epsfig,psfrag}
\usepackage{tabularx}
\usepackage{tikz}
\usepackage{pst-node}
\usepackage{acronym}
\usepackage[yyyymmdd,hhmmss]{datetime}
\usepackage{notation}
\usetikzlibrary{arrows,backgrounds,calc,positioning,shapes,shadows}

\providecommand{\ist}{\hspace*{.3mm}}
\providecommand{\rmv}{\hspace*{-.3mm}}
\providecommand{\iist}{\hspace*{1mm}}
\providecommand{\rrmv}{\hspace*{-1mm}}
\providecommand{\nn}{\nonumber}
\newcommand{\T}{\mathrm{T}}

\acrodef{spa}[SPA]{sum-product algorithm}
\acrodef{da}[DA]{data association}
\acrodef{mmse}[MMSE]{minimum mean-square error}
\acrodef{po}[PO]{potential object}
\acrodef{pmf}[pmf]{probability mass function}
\acrodef{pdf}[pdf]{probability density function}
\acrodef{iid}[iid]{independent and identically distributed}
\acrodef{rmse}[RMSE]{root-mean-squared error}
\acrodef{ospa}[OSPA]{optimal sub-pattern assignment}
\acrodef{bp}[BP]{belief propagation}
\acrodef{bpf}[BPF]{bootstrap particle filter}
\acrodef{upf}[UPF]{unscented particle filter}
\acrodef{pde}[PDE]{partial differential equation}
\acrodef{sde}[SDE]{stochastic differential equation}
\acrodef{ode}[ODE]{ordinary differential equation}
\acrodef{edh}[EDH]{exact Daum and Huang}
\acrodef{ledh}[LEDH]{localized exact Duam and Huang}
\acrodef{pfpf}[PFPF]{particle flow particle filter}
\acrodef{mcmc}[MCMC]{Markov Chain Monte Carlo}
\acrodef{smc}[SMC]{sequential Monte Carlo}
\acrodef{map}[MAP]{maximum a posteriori}
\acrodef{tdoa}[TDOA]{Time-difference of arrival}
\acrodef{pf}[PF]{Particle flow}
\acrodef{mot}[MOT]{multiobject tracking}
\acrodef{pda}[PDA]{probabilistic data association}
\acrodef{jpda}[JPDA]{Joint \ac{pda}}
\acrodef{phd}[PHD]{probability hypothesis density}
\acrodef{cphd}[CPHD]{cardinalized \ac{phd}}
\acrodef{mht}[MHT]{multi-hypothesis tracking}
\acrodef{slam}[SLAM]{simultaneous localization and mapping}
\acrodef{iid}[iid]{independent and identically distributed}
\acrodef{rfs}[RFS]{random finite sets}
\acrodef{ospa}[OSPA]{optimal sub-pattern assignment}
\acrodef{mospa}[MOSPA]{mean \ac{ospa}}

\definecolor{temporalgreen}{RGB}{0,128,0}
\definecolor{spatialred}{RGB}{255,0,0}
\definecolor{temporalblue}{RGB}{0,0,205}

\DeclareMathOperator*{\argmax}{arg\,max}

\allowdisplaybreaks
\providecommand{\bu}{\textcolor{blue}}
\providecommand{\rd}{\textcolor{red}}
\begin{document}
\title{Graph-Based Multiobject Tracking\\ with Embedded Particle \vspace{0mm}Flow}

\author{Wenyu Zhang and Florian Meyer\\[0mm]
University of California San Diego, La Jolla, CA\\[0mm]
Email: \{wez078, flmeyer\}@ucsd.edu
\vspace*{0mm}

\thanks{This material is based upon work supported by the Under Secretary of Defense for Research and Engineering under Air Force Contract No. FA8702-15-D-0001. Any opinions, findings, conclusions or recommendations expressed in this material are those of the authors and do not necessarily reflect the views of the Under Secretary of Defense for Research and Engineering.}
}

\maketitle

\begin{abstract}
Seamless situational awareness provided by modern radar systems relies on effective methods for \ac{mot}. This paper presents a graph-based Bayesian method for nonlinear and high-dimensional \ac{mot} problems that embeds particle flow. To perform operations on the graph effectively, particles are migrated towards regions of high likelihood based on the solution of a partial differential equation. This makes it possible to obtain good object detection and tracking performance with a relatively small number of particles even if object states are high dimensional and sensor measurements are very informative. Simulation results demonstrate reduced computational complexity and memory requirements as well as favorable detection and estimation accuracy in a challenging 3-D \ac{mot} scenario.
\end{abstract}

\begin{IEEEkeywords}
Multiobject tracking, particle flow, factor graphs, sum-product algorithm.
\vspace{-3mm}
\end{IEEEkeywords}

\section{Introduction}\label{sec:introduction}

Multiobject tracking (MOT) is an important capability for a variety of applications including radar surveillance, applied ocean sciences, and autonomous navigation. \ac{mot} is a high-dimensional nonlinear filtering problem complicated by measurement-origin uncertainty and by the fact that the number of the objects to be tracked is \vspace{-2mm} unknown.

\subsection{\ac{mot} and Particle Flow}

Traditional methods for \ac{mot} include \ac{pda} \cite{BarWilTia:B11}, \ac{mht} \cite{Rei:J79}, and methods based on \ac{rfs} \cite{Mah:B07,Wil:J15,SauLiCoa:17}. Most of these traditional approaches suffer from a computational complexity that is combinatorial in both the number of measurements and the number of objects. 
An \ac{mot} method \cite{WilLau:J14,MeyBraWilHla:J17,MeyKroWilLauHlaBraWin:J18} that is based on the framework of factor graphs and the \ac{spa} and is highly scalable in the number of objects, number of measurements, and number of sensors has been proposed recently. This approach uses particle-based computations to calculate messages that, due to nonlinearities in the system model, can not be evaluated in closed form.

As the conventional \acp{bpf} \cite{GorSalSmi:93,AruMasGorCla:02}, existing \ac{spa}-based methods for \ac{mot} \cite{MeyBraWilHla:J17,MeyKroWilLauHlaBraWin:J18} suffer from particle degeneracy \cite{BicLiBen:B08} when object states are high-dimensional or measurements are very informative. This problem is related to the fact that predicted object beliefs are used as proposal distribution for sampling. Since predicted object beliefs often have completely different shapes than the posterior object beliefs needed for state estimation, this sampling strategy is highly inefficient, i.e., few or none of the generated particles are suitable to represent the posterior beliefs. In particular, when object states are high-dimensional or measurements are very informative, sampling from predicted object beliefs fails in the sense that an infeasible large number of particles is needed for accurate state estimation. Thus, since computational complexity and memory requirements are proportional to the number of particles, there is a need for alternative sampling strategies to enable real-time estimation on resource-limited devices.

\ac{pf} \cite{DuaHua:07,DuaHua:09,DuaHua:10,DuaHua:13,BunGod:16} is a promising approach for estimation in nonlinear systems with high-dimensional states or very informative measurements. In \ac{pf}, a homotopy function is used to incrementally migrate a set of particles sampled from a predicted belief such that they finally represent the corresponding posterior object belief. The motion of particles is described by a \ac{pde} that is obtained by combining the homotopy function with the Fokker-Planck equation. In general nonlinear systems, particle flow is suboptimal but can be used for the development of accurate and efficient filtering techniques. In \cite{LiCoates:17}, it is shown that particle flow is an invertible mapping and can thus serve as a proposal distribution in the update step of a particle filter. The resulting \ac{pfpf} is an asymptotically optimal approach to nonlinear filtering that avoids particle degeneracy and provides accurate estimation results even if the number of particles is relatively small. A variant of the \ac{pfpf} has been proposed for \ac{mot}  \cite{SauLiCoa:17}. 


\subsection{Contributions, Paper Organization, and Notation}
We aim to develop a \ac{mot} method that is scalable in relevant system parameters such as number of objects and number of measurements and can succeed also in scenarios with high-dimensional object states and informative measurements. Our approach performs \ac{spa}-based message passing on the factor graph for scalable detection and tracking of an unknown number of objects developed in \cite{MeyKroWilLauHlaBraWin:J18}. To avoid particle degeneracy, we embed invertible particle flow into \ac{spa}-based message passing. This enables migration of particles towards regions of high likelihood and, in turn, leads to an accurate approximation of \ac{spa} messages with a relatively small number of particles. 

In this paper, we introduce a new graph-based \ac{mot} method with invertible particle flow. Contrary to the approach presented in \cite{SauLiCoa:17}, the proposed method performs data association by means of the \ac{spa}. This makes it possible to detect and track a large number of closely spaced objects. Key contributions of this paper are as\vspace{1mm} follows.
\begin{itemize}
\item We establish graph-based \ac{mot} with embedded particle flow that can avoid particle degeneracy in scenarios with high-dimensional object states and informative\vspace{1.5mm} measurements.
\item We demonstrate the reduced computational complexity and favorable detection and estimation accuracy of the proposed method in a challenging 3-D \ac{mot} \vspace{1mm}  scenario.
\end{itemize}

\emph{Notation:} Random variables are displayed in sans serif, upright fonts; their realizations in serif, italic fonts. 
Vectors and matrices are denoted by bold lowercase and uppercase letters, respectively. For example, a random variable and its realization are denoted by $\rv x$ and $x$, respectively, and a random vector and its realization 
by $\RV x$ and $\V x$, respectively. 
Furthermore, $\|\V{x}\|$ and ${\V{x}}^{\text T}$ denote the Euclidean norm and the transpose of vector $\V x$, respectively; and
$\propto$ indicates equality up to a normalization factor. $\Set{N}(\V{x}; \V{x}^{\ast},\M{P})$ denotes the Gaussian \ac{pdf}  (of random vector $\RV{x}$) with mean $\V{x}^{\ast}$ and covariance \vspace{0mm} matrix $\M{P}$.

\section{Review of Invertible Particle Flow}

Let us consider the basic setting of calculating an updated posterior \vspace{0mm} \ac{pdf}
\begin{equation}
  f(\boldsymbol{x}|\boldsymbol{z}) \propto f(\boldsymbol{x}) \ist f(\boldsymbol{z}|\boldsymbol{x}) \nn
  \vspace{.5mm}
\end{equation}
where $\V{x}$ is the state of interest and $\V{z}$ the observed measurement. For the case where the prior \ac{pdf} $f(\boldsymbol{x})$ is Gaussian and the likelihood function $f(\boldsymbol{z}|\boldsymbol{x})$ follows a linear measurement model $\RV{z} = \M{H} \RV{x} + \RV{v}$ with Gaussian measurement noise $\RV{v}$, the posterior \ac{pdf} $f(\boldsymbol{x}|\boldsymbol{z})$ is Gaussian as well and can be calculated in closed form by the update step of the Kalman filter. 

For the case of an arbitrary nonlinear model, e.g., $\RV{z} = \M{h} (\RV{x}) + \RV{v}$, a popular approach is to approximate the posterior $f(\boldsymbol{x}|\boldsymbol{z})$ by weighted samples $\{(\boldsymbol{x}^{(i)}\rmv,w^{(i)})\}^{N_{\mathrm{s}}}_{i=1}$ with $\sum^{N_{\mathrm{s}}}_{i=1} w^{(i)} = 1$, which are calculated based on the importance sampling principle \cite{AruMasGorCla:02}, i.e.\vspace{.5mm},
\begin{equation}
\label{eq:importanceSampling}
  w^{(i)} \propto \frac{f(\boldsymbol{z} |\boldsymbol{x}^{(i)}) \ist f(\boldsymbol{x}^{(i)} )}{q(\boldsymbol{x}^{(i)}|\boldsymbol{z})}.
 \vspace{0mm}
\end{equation}
Here, $q(\boldsymbol{x}|\boldsymbol{z})$ is the proposal \ac{pdf} from which particles $\{\V{x}^{(i)}\}^{N_{\mathrm{s}}}_{i=1}$ are drawn. The \ac{pdf}$q(\boldsymbol{x}|\boldsymbol{z})$ is arbitrary except that it must have the same support as the posterior \ac{pdf} $f(\boldsymbol{x}|\boldsymbol{z})$.

Importance sampling is used in the update step of the conventional particle filter \cite{GorSalSmi:93,AruMasGorCla:02} and is asymptotically optimal for $q(\boldsymbol{x}|\boldsymbol{z})$ ``heavier tailed'', i.e., less informative, than $f(\boldsymbol{x}|\boldsymbol{z})$ \cite{DouFreGor:01}. This means that importance sampling can provide an approximation of $f(\boldsymbol{x}|\boldsymbol{z})$ that can be made arbitrarily good by choosing $N_{\mathrm{s}}$ sufficiently large \cite{AruMasGorCla:02}. For fixed $N_{\mathrm{s}}$, importance sampling is ``more accurate'' if the proposal $q(\boldsymbol{x} |\boldsymbol{z})$ is ``more similar'' to the posterior $f(\boldsymbol{x}|\boldsymbol{z})$ \cite{DouFreGor:01}.

Unfortunately, for conventional  choices \cite{AruMasGorCla:02} of the proposal \ac{pdf} $q(\boldsymbol{x}|\boldsymbol{z})$ and for a feasible number of particles $N_{\mathrm{s}}$, importance sampling often suffers from particle degeneracy \cite{BicLiBen:B08}, especially if the state $\V{x}$ is high-dimensional and/or the measurements $\V{z}$ is very informative (i.e., the likelihood function has sharp and narrow peaks). 

\subsection{\ac{pde} of Particle Flow}

Particle flow \cite{DuaHua:07,DuaHua:09,DuaHua:10,DuaHua:13} is a mechanism that smoothly migrates particles in the state space from the prior \ac{pdf} to the posterior \ac{pdf} by solving a \ac{pde}. Let $\pi(\boldsymbol{x}) \rmv = \rmv  f(\boldsymbol{x}) \ist f(\boldsymbol{z}|\boldsymbol{x})$ be the unnormalized posterior and $l(\boldsymbol{x}) \rmv= \rmv f(\boldsymbol{z}|\boldsymbol{x})$ be the likelihood function. Following \cite{DuaHua:07,DuaHua:09}, a log-homotopy function is introduced \vspace{.5mm} as 
\begin{equation}
  \phi(\boldsymbol{x},\lambda)=\log f(\boldsymbol{x})+\lambda\log l(\boldsymbol{x})
  \label{homotopyPhi}
  \vspace{.5mm}
\end{equation}
where $\lambda \!\in\! [0,1]$ is the pseudo time of the flow process. Note that the homotopy function $\phi(\boldsymbol{x},\lambda) \rmv= \log \pi_\lambda(\boldsymbol{x})$ (with $\pi_\lambda(\boldsymbol{x})\rmv =  f(\boldsymbol{x}) \ist f^\lambda(\boldsymbol{z}|\boldsymbol{x})$) defines the pseudo posterior in the log domain during this flow process, i.e., it defines a continuous and smooth deformation from $\phi(\boldsymbol{x},0) \rmv=\rmv \log f(\boldsymbol{x})$ to $\phi(\boldsymbol{x},1) \rmv= \log \pi(\boldsymbol{x})$. 

Under common assumptions, the stochastic process defined by $\pi_\lambda(\boldsymbol{x})$ (with pseudo time $\lambda$) satisfies the Fokker-Planck equation \cite{BunGod:16,DuaHua:13,DuaHua:10}. By combining the Fokker-Planck for the zero-diffusion case with \eqref{homotopyPhi},  we obtain the following \vspace{.8mm} \ac{pde} \cite{DuaHua:13,DuaHua:10}
\begin{equation}
  \label{exact flow to solve}
      \frac{\partial\phi(\boldsymbol{x},\lambda)}{\partial\boldsymbol{x}}\V{\zeta}(\boldsymbol{x},\lambda)+\log l(\boldsymbol{x})=-\mathrm{Tr}\Big(\frac{\partial \V{\zeta}(\boldsymbol{x},\lambda)}{\partial\boldsymbol{x}}\Big).
       \vspace{1mm}
\end{equation}
Here, $\V{\zeta}(\boldsymbol{x},\lambda) \rmv=\rmv \frac{\mathrm{d} \V{x}}{\mathrm{d}\lambda}$ describes the particle flow, i.e., the migration of particles (samples of $\V{x}$) with pseudo time $\lambda\rmv:\rmv 0\rmv\rightarrow\rmv1$. 

\subsection{Numerical Implementation and Invertible Flow}
\label{sec:numericalImplementation}

For numerical implementation, particle migration is performed by calculating $\V{\zeta}(\boldsymbol{x},\lambda)$ at $N_\lambda$ discrete values of $\lambda$, i.e., $0=\lambda_0<\lambda_1<...<\lambda_{N_\lambda}=1$. Particle flow based on the log-homotopy function  \eqref{homotopyPhi} can be performed as follows. First, $N_\text{s}$ particles $\big\{\V{x}_{0}^{(i)}\big\}_{i=1}^{N_\mathrm{s}}\! =\! \big\{\V{x}_{\lambda_{0}}^{(i)}\}_{i=1}^{N_\mathrm{s}}$ are drawn from $f({\boldsymbol{x}})$. Next, these particles are migrated sequentially across discrete pseudo time steps $l \rmv\in\rmv\{1,\dots,N_{\lambda}\}$,\vspace{1.2mm} i.e.,
\begin{equation}
\boldsymbol{x}^{(i)}_{\lambda_l} = \boldsymbol{x}^{(i)}_{\lambda_{l-1}} + \V{\zeta}(\boldsymbol{x}^{(i)}_{\lambda_{l-1}},\lambda_{l}) (\lambda_{l} - \lambda_{l-1})\hspace{1.5mm}  \label{eq:flow}
\vspace{1mm}
\end{equation}
for all $i\rmv\in\rmv \{1,\dots,N_\mathrm{s}\}$. In this way, particles $\{\V{x}_1^{(i)}\}_{i=1}^{N_\mathrm{s}} \! =\! \{\V{x}_{\lambda_{N_\lambda}}^{(i)}\}_{i=1}^{N_\mathrm{s}}$ representing the posterior \ac{pdf} $\pi(\boldsymbol{x})$ are finally obtained.

If $\log f(\boldsymbol{x})$ and $\log l(\boldsymbol{x})$ are polynomials in the components of the vector $\V{x}$ (e.g., $f(\boldsymbol{x})$ and $l(\boldsymbol{x})$ are Gaussians or in another exponential family), \eqref{exact flow to solve} can be solved exactly and in closed form. This closed-form flow solution is used in the update step of the \ac{edh} filter \cite{DuaHua:10Sol,DuaHua:10}.
In particular, let us consider a Gaussian prior $f(\V{x}) = \Set{N}(\V{x}; \V{x}_0^{\ast},\M{P})$ with mean ${\boldsymbol{x}}_0^\ast$ and covariance matrix $\M{P}$ as well as a linear measurement model $\RV{z} = \M{H} \RV{x} + \RV{v}$. Here, the measurement noise $\RV{v}$ is zero-mean Gaussian with covariance matrix $\M{R}$. The exact flow solution \cite{DuaHua:10Sol,DuaHua:10} is now given \vspace{1.2mm} by 
\begin{equation}
  \V{\zeta}(\boldsymbol{x},\lambda) = \M{A}(\lambda)\boldsymbol{x} + \V{b}(\lambda)  \nn
  \vspace{0mm}
\end{equation}
where
\vspace{1mm}
\begin{align}
  \M{A}(\lambda) &= -\frac{1}{2}\M{P}\M{H}^\T(\lambda \M{H}\M{P}\M{H}^\T+\M{R})^{-1}\M{H} \label{EDH_A}\\[1mm]
  \V{b}(\lambda) &= (\M{I}+2\lambda \M{A}(\lambda))\big[(\M{I}+\lambda \M{A}(\lambda))\M{P}\M{H}^\T \M{R}^{-1}\boldsymbol{z}+\M{A}(\lambda){\boldsymbol{x}}^\ast_0\big].  \nn\\[-2.5mm] 
  \nn
\end{align}
For nonlinear measurement models $\RV{z} = \M{h} (\RV{x}) + \RV{v}$, a suboptimal linearization step is employed. In particular, at each step $l \rmv\in\rmv\{1,\dots,N_{\lambda}\}$, a first-order Taylor series approximation is performed to calculate an approximate measurement matrix $\tilde{\M{H}}_{\lambda_{l-1}}$ from $\M{h}(\cdot)$  \vspace{-.3mm} at a current mean ${\V{x}}^\ast_{\lambda_{l-1}}$. This mean is propagated in parallel to the particles $\{\V{x}_{\lambda_{l-1}}^{(i)}\}_{i=1}^{N_\mathrm{s}}$ by using \eqref{eq:flow} (see \cite[Section~III-A]{LiCoates:17} for details).
Particle flow based on this linearized model has no optimality guarantees but has been demonstrated numerically to typically provide an accurate representation of $f(\V{x}|\V{z}).$

For asymptotically optimal estimation, particle flow can be used as proposal \ac{pdf} $q(\boldsymbol{x}|\boldsymbol{z})$ for importance sampling (cf. \eqref{eq:importanceSampling}). In particular, the particle flow mapping $\RV{x}_{0}\rmv\rightarrow\rmv \RV{x}_{1}$ is proven to be invertible, i.e., under certain constraints on the differences of consecutive discrete pseudo times $\lambda_l - \lambda_{l-1}$, $l \rmv\in\rmv\{1,\dots,N_{\lambda}\}$ \cite{LiCoates:17}, there exists a mapping of the particles at $\lambda_{N_{\lambda}} = 1$ to the particles at $\lambda_0 = 0$. By exploiting this invertible mapping, the proposal \ac{pdf} related to particle flow can be evaluated\vspace{1.5mm} as
\begin{equation}
  q(\boldsymbol{x}_{1}^{{(i)}}|\boldsymbol{z}) = \frac{f(\V{x}_{0}^{(i)})} {\theta}
  \label{eq:invertibleMapping}
  \vspace{2mm}
\end{equation}
where the ``mapping factor'' $\theta$ is given\vspace{1.5mm} by
\begin{equation}
\theta = \prod_{l=1}^{N_\lambda} \big| \ist \mathrm{det}\big[\V{I}+(\lambda_{l} - \lambda_{l-1}) \ist \tilde{\M{A}}(\lambda_l)\big] \big|. 
\label{eq:mappingFactor}
\vspace{1.5mm}
\end{equation}
Here, $\tilde{\M{A}}(\lambda_l)$ is the approximation of $\M{A}(\lambda)$ in \eqref{EDH_A} at $\lambda_l$ based on the first-order Taylor series approximation of the measurement model discussed above. Performing importance sampling in \eqref{eq:importanceSampling} by using flow particles  $\{\V{x}_1^{(i)}\}_{i=1}^{N_\mathrm{s}}$ and by evaluating the corresponding proposal \ac{pdf}  in \eqref{eq:invertibleMapping} is asymptotically optimal and can provide accurate estimation results in challenging nonlinear and high-dimensional estimation problems even with a moderate number of particles\vspace{-3mm}  \cite{LiCoates:17}.

\section{Graph-based Multiobject Tracking}
\label{sec:review} \vspace{-1mm}
In what follows, we will review the system model for graph-based \ac{mot} presented in \cite{MeyKroWilLauHlaBraWin:J18}.

\subsection{System Model}\label{sec:sceass}
\subsubsection{Potential Object States and State-Transition Function} 
As in \cite{MeyBraWilHla:J17,MeyKroWilLauHlaBraWin:J18}, we consider \ac{mot} for an unknown, time-varying number of objects by introducing \acp{po}. The number of \acp{po} $j_k$ at discrete time $k \rmv\geq\rmv 0$ is the maximum possible number of objects that have generated a measurement so far.
We introduce the augmented state for a \ac{po} $j \in \{1,\dots,j_k\}$ as $\RV{y}_{k}^{(j)} \rmv\triangleq\rmv \big[\RV{x}^{(j)\ist\T}_{k} \; \rv{r}^{(j)}_{k} \big]^\T\rmv\rmv\rmv$.
Here, the existence variable $\rv{r}^{(j)}_{k} \rmv\in\rmv \{0,1\}$ models the existence/nonexistence of \ac{po} $j$ in the sense that \ac{po} $j$ exists at time $k$ if and only if $\rv{r}^{(j)}_{k} \!=\! 1$.
The state $\RV{x}^{(j)}_{k}$ of \ac{po} $j$ consists of the \ac{po}'s position and possibly further parameters. The state $\RV{x}^{(j)}_{k}$ of nonexistent \acp{po} are obviously irrelevant. Therefore, all \acp{pdf} defined for augmented \acp{po}' states, $f\big(\V{y}^{(j)}_{k}\big) =\rmv f\big(\V{x}^{(j)}_{k}\rmv, r^{(j)}_{k}\big)$, have the property that $f\big(\V{x}^{(j)}_{k}\rmv, 0 \big) = f^{(j)}_{k} f_{\text{D}}\big(\V{x}^{(j)}_{k}\big)$, where $f_{\text{D}}\big(\V{x}^{(j)}_{k}\big)$ is an arbitrary ``dummy \ac{pdf}'' and $f^{(j)}_{k} \!\rmv\in [0,1]$ is a constant.
For each \ac{po} state $\RV{y}_{k-1}^{(j)}$, $j \in \{1,\dots,j_{k-1}\}$ at time $k-1$, there is one ``legacy'' \ac{po} state $\underline{\RV{y}}_{k}^{(j)}$ at time $k$. It is assumed that each object evolves independently in time. The single-object state transition \ac{pdf}  $f\big( \underline{\V{y}}_k^{(j)} \big| \V{y}_{k-1}^{(j)} \big)$ that models the motion and disappearance of objects and involves the probability of object survival $p_{\text{s}}$ is presented in \cite[Section VIII-C]{MeyKroWilLauHlaBraWin:J18}. At time $k \rmv=\rmv 0$, the prior augmented states $\V{y}^{(j)}_{0}$ are statistically independent across \acp{po} $j$. Often no prior information is available, i.e., $j_0 \rmv=\rmv 0$.
\vspace{1.5mm}

\subsubsection{New \acp{po}, Data Association, and Measurement Likelihood Function}
A sensor produces measurements $\RV{z}^{(m)}_{k} \rmv$, $m \rmv\in\rmv \big\{ 1,\dots,\rv{m}_k \big\}$ at each time $k\rmv\geq\rmv1$. (Note that the number of measurements $\rv{m}_k$ is random.) Each measurement can originate from one of the three sources: (i) a legacy \ac{po}, which represents an object that has generated at least one measurement before; (ii) a new \ac{po}, which models an object that generates a measurement for the first time;  and (iii) clutter. The birth of new objects is modeled by a Poisson point process with mean $\mu_{\text{b}}$ and \ac{pdf} $f_{\text{b}}\big( \overline{\V{x}}_{k} \big)$. At time $k$, $\rv{m}_k$ new \ac{po} states are introduced, i.e., $\overline{\RV{y}}_{k}^{(m)}\rmv\triangleq\rmv \big[\overline{\RV{x}}^{(m)\ist\T}_{k} \; \overline{\rv{r}}^{(m)}_{k} \big]^\T\rmv\rmv$, $m \in \{1,\dots,\rv{m}_k\}$. Here, $\overline{\rv{r}}^{(m)}_{k} = 1$ means that measurement $\RV{z}^{(m)}_{k}$ originated from an object that never generated a measurement before, and $\overline{\rv{r}}^{(m)}_{k} = 0$ otherwise. After the observation, the joint measurement vector at time $k$ is fixed and denoted as $\RV{z}_k \!\triangleq\rmv \big[\RV{z}_{k}^{(1)\ist\T} \rmv\cdots\ist \RV{z}_{k}^{(m_k)\ist\T}\big]^{\T}\rmv\rmv.$ The total number of (legacy and new) \acp{po} states at time $k$ is $j_{k} \!=\rmv j_{k-1} \!+\rmv m_{k}$ and the joint \ac{po} state at time $k$ is denoted as $\RV{y}_k \!\triangleq\rmv \big[\RV{y}_{k}^{(1)\ist\T} \rmv\cdots\ist \RV{y}_{k}^{(j_k)\ist\T}\big]^{\T}\rmv\rmv.$

The object represented by \ac{po} $j \in \{1,\dots,j_k\}$ is detected (in the sense that it generates a measurement $\RV{z}_k^{(m)}$) with probability $p_{\text{d}}$. The statistical relationship of a measurement $\RV{z}_k^{(m)}$ and a detected \ac{po} state $\RV{x}_k^{(j)}$ is described by the conditional \ac{pdf} $f\big(\V{z}^{(m)}_{k} | \V{x}^{(j)}_{k} \big)$, which is based on the measurement model of the sensor. Clutter measurements are modeled by a Poisson point process with mean $\mu_{\text{c}}$ and \ac{pdf} $f_{\text{c}}\big( \V{z}^{(m)}_{k} \big)$.

In \ac{mot}, measurements are subject to \ac{da} uncertainty: it is unknown which measurement originated from which \ac{po}, and a measurement may also be clutter, i.e., originating from any \ac{po}.
We make the point object assumption, i.e., at any time $k$, an object can generate at most one measurement and a measurement can originate from at most one object \cite{BarWilTia:B11,Mah:B07,MeyKroWilLauHlaBraWin:J18}.
Then the association between $m_k$ measurements and $j_{k-1}$ legacy \acp{po} at time $k$ can be modeled by an ``object-oriented'' \ac{da} vector $\RV{a}_{k} = \big[\rv{a}_{k}^{(1)} \cdots\ist \rv{a}_{k}^{(j_{k-1})} \big]^{\T}\rmv\rmv$. The object-oriented association variable $\rv{a}_{k}^{(j)}$ is $m \in \{1,\dots,m_{k} \}$ if \ac{po} $j$ generates measurement $m$ and zero if \ac{po} $j$ is missed by the sensor \cite{BarWilTia:B11,MeyKroWilLauHlaBraWin:J18}. We also introduce the ``measurement-oriented'' \ac{da} vector $\RV{b}_{k} = \big[\rv{b}_{k}^{(1)} \cdots\ist \rv{b}_{k}^{(m_k)} \big]^{\T}\rmv\rmv$ to obtain a scalable and efficient message passing algorithm (see  \cite{WilLau:J14,MeyKroWilLauHlaBraWin:J18} for details). The measurement-oriented association variable $\rv{b}_{k}^{(m)}$ is $j \in \{1,\dots,j_{k-1} \}$ if measurement $m$ originated from legacy \ac{po} $j$ and zero if it originated from clutter or a newly detected \vspace{1.5mm}\ac{po}.

\subsubsection{Joint Posterior PDF and Factor Graph}
\label{eq:JointAssociation}

Using common assumptions \cite{BarWilTia:B11,Rei:J79,Mah:B07,Wil:J15,WilLau:J14,MeyBraWilHla:J17,MeyKroWilLauHlaBraWin:J18}, the joint posterior \ac{pdf} of $\RV{y}_{1:k}$, $\RV{a}_{1:k}$, and $\RV{b}_{1:k}$ conditioned on observed and thus fixed $\V{z}_{1:k}$ can be obtained \vspace{.5mm}as
\begin{align}
  &f\big( \V{y}_{1:k}, \V{a}_{1:k}, \V{b}_{1:k} \big| \V{z}_{1:k} \big) \nn \\[0.5mm]
  &\hspace{0mm} \propto  \Bigg(\prod^{j_{0}}_{j''=1} f\big(\V{y}^{(j'')}_{0} \big)  \Bigg)  \prod^{k}_{k'=1}   \Bigg(\prod^{j_{k'-1}}_{j'=1} f\big(\underline{\V{y}}^{(j')}_{k'}\big|\V{y}^{(j')}_{k'-1}\big)  \Bigg)   \nn\\[0.5mm] 
  &\hspace{4.5mm}\times \Bigg( \prod^{j_{k'-1}}_{j=1}  q\big( \underline{\V{x}}^{(j)}_{k'}\!, \underline{r}^{(j)}_{k'}\!\rmv, a^{(j)}_{k'}\rmv; \V{z}_{k'} \big)\rmv\rmv \prod^{m_{k'}}_{m'=1} \Psi_{j\rmv,m'}\big(a_{k'}^{(j)}\rmv,b_{k'}^{(m')}\big) \rmv\Bigg)  \nn\\[1.5mm]
  &\hspace{4.5mm}\times  \hspace{1mm} \prod^{m_{k'}}_{m=1} \rmv v\big( \overline{\V{x}}^{(m)}_{k'}\!, \overline{r}^{(m)}_{k'}\!, b^{(m)}_{k'}\rmv; \V{z}_{k'}^{(m)} \big)\hspace{-.1mm}.
  \label{eq:jointPosteriorComplete} \\[-6mm]
  \nn
\end{align}
Here, the legacy \ac{po} pseudo likelihood function $q\big( \underline{\V{x}}^{(j)}_{k}\!, \underline{r}^{(j)}_{k},$ $a^{(j)}_{k};\V{z}_{k} \big)$ is given by
\begin{align}
&\hspace{-.8mm}q\big( \underline{\V{x}}^{(j)}_{k}\!, 1, a^{(j)}_{k}\rmv; \V{z}_{k} \big) \nn\\
&\hspace{-.8mm}\triangleq \begin{cases}
    \frac{ p_{\text{d}} }{ \mu_{\text{c}} f_{\text{c}}\big( \V{z}_{k}^{(m)} \big)} f\big( \V{z}_{k}^{(m)} \rmv\big|\ist \underline{\V{x}}_{k}^{(j)} \big), & a^{(j)}_{k} \rmv= m \rmv\in\rmv \{1,\dots,m_k \}  \\[2mm]
     1 - p_{\text{d}}  \ist, & a^{(j)}_{k} \rmv=\rmv 0 \ist
  \end{cases}
  \nn\\[-5.5mm]
  \nn
\end{align}
and $q\big( \underline{\V{x}}^{(j)}_{k}\!, 0, a^{(j)}_{k}\rmv; \V{z}_{k} \big) \triangleq 1(a^{(j)}_{k})$, where $1(a)$ denotes the indicator function of the event $a \rmv=\rmv 0$ (i.e., $1(a) \rmv=\rmv 1$ if $a \rmv=\rmv 0$ and $0$ otherwise). Furthermore, the new \ac{po} pseudo likelihood function $v\big( \overline{\V{x}}^{(m)}_{k}\!, \overline{r}^{(m)}_{k},$ $b^{(m)}_{k}\rmv; \V{z}_{k}^{(m)} \big)$ reads
\begin{align}
&\hspace{-.9mm}v\big( \overline{\V{x}}^{(m)}_{k}\!, 1, b^{(m)}_{k}\rmv; \V{z}_{k}^{(m)} \big) \nn \\[.5mm]
&\hspace{-.9mm}\triangleq \begin{cases}
     0 \ist,  & \hspace{-1mm} b^{(m)}_{k} \rmv\in\rmv \{ 1,\dots,j_{k-1} \}\\[2mm]
     { \ist \frac{p_{\mathrm{d}} \ist \mu_{\text{b}} \ist f_{\text{b}}\big(\overline{\V{x}}^{(m)}_{k}\big)}{ \mu_{\text{c}} f_{\text{c}}\big( \V{z}_{k}^{(m)} \big)} }  f\big(\V{z}_k^{(m)}  \big| \overline{\V{x}}^{(m)}_{k} \big) \ist, & \hspace{-1mm} b^{(m)}_{k} \!=\rmv 0 
  \end{cases}
  \nn\\[-4.0mm]
  \nn
\end{align}
and $v\big( \overline{\V{x}}^{(m)}_{k}\!, 0, b^{(m)}_{k}\rmv; \V{z}_{k}^{(m)} \big) \rmv\triangleq\rmv f_{\text{D}}\big(\overline{\V{x}}^{(m)}_{k}\big)$.

Finally, the binary indicator function $\Psi_{j\rmv,m}\big(a_{k}^{(j)}\rmv,b_{k}^{(m)}\big)$ checks association consistency of a pair of object-oriented and measurement-oriented variables $\big(a_{k}^{(j)}\rmv,b_{k}^{(m)}\big)$ in that $\Psi_{j\rmv,m}\big(a_{k}^{(j)}\rmv,b_{k}^{(m)}\big)$ is zero if 
$a_{k}^{(j)} \rmv= m, b^{(m)}_{k} \rmv\neq\rmv j$ or $b^{(m)}_{k} \rmv=\rmv j, a_{k}^{(j)} \rmv\neq\rmv m$ and one otherwise (see \cite{WilLau:J14,MeyKroWilLauHlaBraWin:J18} for details). The joint posterior in \eqref{eq:jointPosteriorComplete} can be represented by the factor graph in Fig.~\ref{fig:factorGraphU}. A detailed derivation of this joint posterior is provided in \cite[Section~VIII-G]{MeyKroWilLauHlaBraWin:J18}. 

\subsection{Problem Formulation and Selected Message Passing Operations}

We consider the problem of object detection and state estimation at time $k \rmv\geq\rmv 1$ based on all measurements $\V{z}_{1:k}$ collected up to time $k$. Object detection is performed by comparing the existence probability $p\big(r_{k}^{(j)}\! \rmv=\rmv 1 \big| \V{z}_{1:k} \big)$ with a threshold $P_{\text{th}}$, i.e., \ac{po} $j \in \{1,\dots,j_k\}$ is declared to exist if $p\big(r_{k}^{(j)}\! \rmv=\rmv 1 \big| \V{z}_{1:k} \big) \rmv>\rmv P_{\text{th}}$. Note that $p\big(r_{k}^{(j)}\! \rmv=\rmv 1 \big| \V{z}_{1:k} \big) \rmv= \int f\big(\V{x}_k^{(j)}, r_k^{(j)}\! \rmv=\rmv 1 \big| \V{z}_{1:k}\big) \ist\mathrm{d}\V{x}_k^{(j)}$.
For existent \acp{po}, state estimation is performed by calculating the \ac{mmse} estimate \cite{Poo:B94}  as
\vspace{0mm}
\begin{equation}
\hat{\V{x}}_k^{(j)}
\ist\triangleq \int \V{x}_k^{(j)} f\big(\V{x}_k^{(j)} \big| r_k^{(j)} \rmv=\rmv 1, \V{z}_{1:k}\big) \ist\mathrm{d}\V{x}_k^{(j)}
\label{eq:mmseEst}
\end{equation} 
where $f\big(\V{x}_k^{(j)} \big| r_k^{(j)} \rmv=\rmv 1, \V{z}_{1:k}\big) \rmv=\rmv f\big(\V{x}_k^{(j)}, r_k^{(j)} \rmv=\rmv 1 \big| \V{z}_{1:k}\big)/ p\big(r_{k}^{(j)}$ $=\rmv 1\big| \V{z}_{1:k} \big)$.

\begin{figure}[t!]
\centering
\psfrag{A1}[l][l][0.8]{}
\psfrag{A2}[l][l][0.8]{}

\psfrag{B1}[l][l][0.8]{\raisebox{-2.4mm}{\hspace{-.5mm}$\bu{\alpha_1}$}}
\psfrag{B2}[l][l][0.8]{\raisebox{-4.4mm}{\hspace{-1.8mm}$\bu{\alpha_{n_{\mathrm{p}}}}$}}

\psfrag{C1}[l][l][0.8]{\raisebox{-16mm}{\hspace{-6mm}$\rd{\gamma_1}$}}
\psfrag{C2}[l][l][0.8]{\raisebox{-3.5mm}{\hspace{-1.2mm}$\rd{\gamma_{n_{\mathrm{p}}}}$}}

\psfrag{D1}[l][l][0.8]{\raisebox{1.5mm}{\hspace{.1mm}$\bu{\alpha_1}$}}
\psfrag{D2}[l][l][0.8]{\raisebox{-1.7mm}{\hspace{-1mm}$\bu{\alpha_{n_{\mathrm{p}}}}$}}

\psfrag{F1}[l][l][0.8]{\raisebox{-4mm}{\hspace{1mm}$\rd{\beta_1}$}}
\psfrag{F2}[l][l][0.8]{\raisebox{-4mm}{\hspace{.8mm}$\rd{\beta_{n_{\mathrm{p}}}}$}}

\psfrag{E1}[l][l][0.8]{\raisebox{-2.2mm}{\hspace{0mm}$\bu{\kappa_1}$}}
\psfrag{E2}[l][l][0.8]{\raisebox{-2.6mm}{\hspace{.1mm}$\bu{\kappa_{n_{\mathrm{p}}}}$}}

\psfrag{H1}[l][l][0.8]{}
\psfrag{G1}[l][l][0.8]{}
\psfrag{H2}[l][l][0.8]{}
\psfrag{G2}[l][l][0.8]{}
\psfrag{I1}[l][l][0.8]{}
\psfrag{J1}[l][l][0.8]{}
\psfrag{I2}[l][l][0.8]{}
\psfrag{J2}[l][l][0.8]{}

\psfrag{K1}[l][l][0.8]{\raisebox{-3.0mm}{\hspace{-.3mm}$\rd{\xi_{1}}$}}
\psfrag{K2}[l][l][0.8]{\raisebox{-2.5mm}{\hspace{-.5mm}$\rd{\xi_{n_{\mathrm{m}}}}$}}

\psfrag{L1}[l][l][0.8]{\raisebox{-1.0mm}{\hspace{.7mm}$\bu{\iota_{1}}$}}
\psfrag{L2}[l][l][0.8]{\raisebox{1.9mm}{\hspace{-1.2mm}$\bu{\iota_{n_{\mathrm{m}}}}$}}

\psfrag{M1}[l][l][0.8]{\raisebox{1mm}{\hspace{1.7mm}$\rd{\varsigma_{1}}$}}
\psfrag{M2}[l][l][0.8]{\raisebox{0.5mm}{\hspace{-.5mm}$\rd{\varsigma_{n_{\mathrm{m}}}}$}}

\psfrag{c1}[l][l][.75]{\raisebox{-3.5mm}{\hspace{0mm}$\underline{\V{y}}^{1}$}}
\psfrag{c2}[l][l][.75]{\raisebox{-2.8mm}{\hspace{-.4mm}$\underline{\V{y}}^{n_{\mathrm{p}}}$}}

\psfrag{d1a}[l][l][.75]{\raisebox{-3mm}{\hspace{1.5mm}$f^{1}$}}
\psfrag{d2a}[l][l][.75]{\raisebox{-3mm}{\hspace{.7mm}$f^{n_{\mathrm{p}}}$}}

\psfrag{c1a}[l][l][.73]{\raisebox{-3mm}{\hspace{1.3mm}$\overline{\V{y}}^{\ist 1}$}}
\psfrag{c2a}[l][l][.73]{\raisebox{-3.2mm}{\hspace{0mm}$\overline{\V{y}}^{\ist n_{\mathrm{m}}}$}}

\psfrag{f1a}[l][l][.75]{\raisebox{-0.8mm}{\hspace{.4mm}$q^1$}}
\psfrag{f3a}[l][l][.75]{\raisebox{-3.0mm}{\hspace{-1.4mm}$\;q^{n_{\mathrm{p}}}$}}

\psfrag{f1b}[l][l][.75]{\raisebox{.3mm}{\hspace{.7mm}$v^1$}}
\psfrag{f3b}[l][l][.75]{\raisebox{-2.1mm}{\hspace{-.3mm}$v^{n_{\mathrm{m}}}$}}

\psfrag{a1}[l][l][.7]{\raisebox{-1.8mm}{\hspace{.4mm}$a^{1}$}}
\psfrag{a2}[l][l][.7]{\raisebox{-1.6mm}{\hspace{0mm}$a^{n_{\mathrm{p}}}$}}

\psfrag{b1}[l][l][.7]{\raisebox{-2.4mm}{\hspace{.2mm}$b^{1}$}}
\psfrag{b4}[l][l][.7]{\raisebox{-2mm}{\hspace{-.4mm}$b^{n_{\mathrm{m}}}$}}

\psfrag{psi1}[l][l][.7]{\raisebox{-4.0mm}{\hspace{-1mm}$\Psi^{1,1}$}}
\psfrag{psi2}[l][l][.7]{\raisebox{-4mm}{\hspace{-2mm}$\Psi^{n_{\mathrm{p}},n_{\mathrm{m}}}$}}
\psfrag{psi3}[l][l][.7]{\raisebox{-3.5mm}{\hspace{-2.5mm}$\Psi^{1,n_{\mathrm{m}}}$}}
\psfrag{psi4}[l][l][.7]{\raisebox{-2mm}{\hspace{-2.3mm}$\Psi^{n_{\mathrm{p}},1}$}}

\hspace{1mm}\includegraphics[scale=.7]{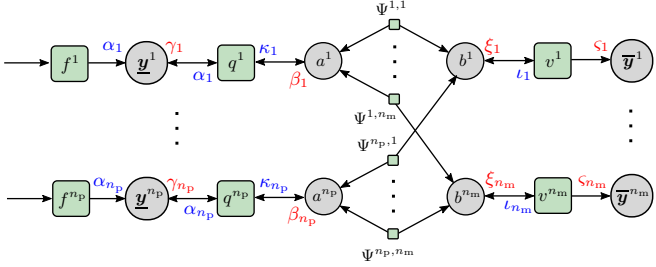}

  \caption{Factor graph for \ac{mot} of an unknown, time-varying number of objects, corresponding to the propagation of the joint \ac{pdf} $f\big( \V{y}_{1:k}, \V{a}_{1:k}, \V{b}_{1:k} \big| \V{z}_{1:k} \big)$ in \eqref{eq:jointPosteriorComplete}. One time step $k$ is shown. Messages calculated using particle flow are depicted in \textcolor{spatialred}{red}. These messages are calculated based on messages depicted in \textcolor{blue}{blue}.
 The time index $k$ is omitted, and the following short notations are used: 
$n_{\mathrm{m}} \rmv\triangleq m_{k}$, 
$n_{\mathrm{p}} \rmv\triangleq j_{k-1}$,
$\underline{\V{y}}^{j} \rmv\triangleq \underline{\V{y}}_{k}^{(j)}$,
$\overline{\V{y}}^{\ist m} \rmv\triangleq \overline{\V{y}}_{k}^{(m)}\rmv$,
$\V{a}^j \rmv\triangleq \V{a}^{(j)}_k$,
$\V{b}^m \rmv\triangleq \V{b}^{(m)}_k$,
\textcolor{temporalgreen}{$f^j \rmv\triangleq f\big(\underline{\V{y}}^{(j)}_{k} \big| \V{y}^{(j)}_{k-1}\big)$},
 \textcolor{temporalgreen}{$q^j \rmv\triangleq q\big( \underline{\V{x}}^{(j)}_{k}\rmv, \underline{r}^{(j)}_{k}\rmv, a_{k}^{(j)};\V{z}_{k} \big)$}, 
 \textcolor{temporalgreen}{$v^m \rmv\triangleq v\big( \overline{\V{x}}^{(m)}_{k} \rmv,\overline{r}^{(m)}_{k} \rmv,b_{k}^{(m)};\V{z}^{(m)}_{k} \big)$},
\textcolor{temporalgreen}{$\Psi^{j,m} \rmv\triangleq \Psi_{j,m}\big(a^{(j)}_{k} \rmv,b^{(m)}_{k}\big)$}, \textcolor{red}{$\gamma_j \triangleq \gamma_{k}^{(j)}\big(\underline{\V{y}}^{(j)}_{k} \big)$}, \textcolor{red}{$\beta_j \triangleq \beta^{(j)}_{k}\big(a^{(j)}_{k}\big)$},  \textcolor{red}{$\xi_m \triangleq \xi^{(m)}_{k}\big(b^{(m)}_{k}\big)$}, \textcolor{red}{$\varsigma_{m} \triangleq \varsigma^{(m)}_k\big(\overline{\V{y}}^{(m)}_{k} \big)$}, \textcolor{blue}{$\alpha_j \triangleq \alpha_k\big(\underline{\V{y}}^{(j)}_{k}\big)$}, \textcolor{blue}{$\kappa_j \triangleq \kappa^{(j)}_{k}\big(a^{(j)}_{k} \big)$}, and \textcolor{blue}{$\iota_{m} \triangleq \iota_k^{(m)}\big(b^{(m)}_{k} \big)$}.
\label{fig:factorGraphU}
\vspace{0mm}
}
\end{figure}

Both object detection and estimation require the marginal posterior \acp{pdf} $f\big(\V{x}_k^{(j)}, r_k^{(j)} \big| \V{z}_{1:k})\rmv \triangleq f\big(\V{y}_k^{(j)} \big| \V{z}_{1:k})$, $j \in \{1,\dots,j_k\}$. However, calculating $f\big(\V{x}_k^{(j)}, r_k^{(j)} \big| \V{z}_{1:k}\big)$ by direct marginalization of (\ref{eq:jointPosteriorComplete}) is infeasible due to the high dimensionality of $\V{y}_{1:k}$, $\V{a}_{1:k}$, and $\V{b}_{1:k}$. As in \cite{MeyBraWilHla:J17,MeyKroWilLauHlaBraWin:J18}, we consider approximate calculation by performing the loopy \ac{spa} on the factor graph in Fig.~\ref{fig:factorGraphU} and passing messages only forward in time. This makes it possible to efficiently calculate so-called beliefs $\tilde{f}\big(\V{x}_k^{(j)}\!, r_k^{(j)}\big), \ist j \in \{1,\dots,j_{k} \}$ which accurately approximate the marginal posterior \acp{pdf} $f\big(\V{x}_k^{(j)}\!, r_k^{(j)} \big| \V{z}_{1:k}\big), \ist j \in \{1,\dots,j_{k} \}$ needed for object detection and estimation. To keep computational complexity feasible, at each time $k$, a suboptimal pruning step has to be performed.  In particular, at each time $k$, \acp{po} with probability of existence $p(r^{(j)}_{k} \!=\! 1|\V{z}^{(1:n)})$ below a threshold $P_{\text{pr}}$ are removed from the state space.

Next, we present the \ac{spa} messages that will later be calculated based on particle flow. We will limit our discussion to messages and beliefs related to legacy \ac{po} states. Messages and beliefs related to new \ac{po} states are obtained by performing similar steps. (A complete description of message passing for \ac{mot} is provided in \cite[Section~IX-A]{MeyKroWilLauHlaBraWin:J18}.) Messages calculated by particle flow are highlighted in the factor graph in Fig.~\ref{fig:factorGraphU}.  After calculating messages $\alpha_k^{(j)}\big(\underline{\V{x}}_k^{(j)},\underline{r}_k^{(j)}\big)$, $j \rmv\in\rmv \{1,\dots,j_{k-1}\}$  in a ``prediction'' step \cite[Section~IX-A1]{MeyKroWilLauHlaBraWin:J18}, a ``measurement evaluation'' step is performed. For future reference, we also introduce $\alpha_{\mathrm{e},k}^{(j)} \rmv=\rmv \int \alpha_k^{(j)}\big(\underline{\V{x}}_k^{(j)},\underline{r}_k^{(j)} \rmv=\rmv1\big) \mathrm{d} \underline{\V{x}}_k^{(j)}$ and $\alpha_{\mathrm{n},k}^{(j)} \rmv=\rmv \int \alpha_k^{(j)}\big(\underline{\V{x}}_k^{(j)}\rmv\rmv,\underline{r}_k^{(j)} \rmv=\rmv0\big) \mathrm{d} \underline{\V{x}}_k^{(j)}$.

For legacy \acp{po}, the messages $\beta_k^{(j)}\big(a_k^{(j)}\big)$ passed from factor nodes $q\big( \underline{\V{x}}^{(j)}_{k}\!, \underline{r}^{(j)}_{k}\!, a^{(j)}_{k}\rmv; \V{z}_{k} \big)$ to variable nodes $a^{(j)}_{k}$ are calculated as
\begin{align}
  &\beta_k^{(j)}\big(a_k^{(j)}\big) = \rmv\int\rmv q\big( \underline{\V{x}}^{(j)}_{k}\!, 1, a^{(j)}_{k}\rmv; \V{z}_{k} \big) \alpha_k^{(j)}\!\big(\underline{\V{x}}_k^{(j)}\!,1\big) \mathrm{d}\underline{\V{x}}_k^{(j)} \nn\\
  &\hspace{52.5mm}+ 1\big(a_k^{(j)}\big)\alpha_{\mathrm{n},k}^{(j)} \ist.
  \label{eq:messageBeta}
\end{align}
For new \acp{po}, messages $\xi_k^{(m)}\big(b_k^{(m)}\big)$ are calculated similarly (see \cite[Section~IX]{MeyKroWilLauHlaBraWin:J18}).
Now, probabilistic \ac{da} is performed by using the iterative \ac{spa}-based algorithm \cite[Section~IX-A3]{MeyKroWilLauHlaBraWin:J18} with input messages $\beta_k^{(j)}\big(a_k^{(j)}\big)$, $j \in \{1,\dots,j_k\}$ and $\xi_k^{(m)}\big(b_k^{(m)}\big)$, $m \in \{1,\dots,m_k\}$. After convergence, corresponding output messages $\kappa_k^{(j)}\big(a_k^{(j)}\big)$, $j \in \{1,\dots,j_k\}$ and $\iota_k^{(m)}\big(b^{(m)}_{k} \big)$, $m \in \{1,\dots,m_k\}$ are available for legacy \acp{po} and new \acp{po}, respectively. 

Next, a ``measurement update'' step is performed. For legacy \acp{po}, messages $\gamma_k^{(j)}\big(\underline{\V{x}}_k^{(j)}\!, \underline{r}_k^{(j)}\big)$ passed from $q\big( \underline{\V{x}}^{(j)}_{k}\!, \underline{r}^{(j)}_{k}\!, a^{(j)}_{k}\rmv; \V{z}_{k} \big)$ to $\underline{\V{y}}_k^{(j)}$ are calculated \vspace{0mm}as
\begin{equation}
  \gamma_k^{(j)}\big(\underline{\V{x}}_k^{(j)}\!, 1\big) = \rmv\sum_{a_k^{(j)}\rmv=0}^{m_k} q\big( \underline{\V{x}}^{(j)}_{k}\!, 1, a^{(j)}_{k}\rmv; \V{z}_{k} \big)  \kappa_k^{(j)}\big(a_k^{(j)}\big)
  \label{eq:messageGamma}
  \vspace{0mm}
\end{equation}
and $\gamma_k^{(j)}\big(\underline{\V{x}}_k^{(j)}\!, 0\big) = \gamma_k^{(j)} =  \kappa_k^{(j)}\big(0\big)$.
Finally, beliefs are calculated to approximate the posterior \acp{pdf} of \acp{po}. In particular, for legacy \acp{po}, beliefs $\tilde{f}\big(\underline{\V{x}}_k^{(j)}\!, \underline{r}_k^{(j)}\big)$ approximating $f\big(\underline{\V{x}}_k^{(j)}\!, \underline{r}_k^{(j)}\rmv\big|\rmv\V{z}_{1:k}\big)$ are obtained\vspace{.5mm} as 
\begin{equation}
  \tilde{f}\big(\underline{\V{x}}_k^{(j)}\!, 1\big) \rmv=\rmv \frac{1}{\underline{C}_k^{(j)}}\alpha_k^{(j)}\big(\underline{\V{x}}_k^{(j)}\!, 1\big)\gamma_k^{(j)}\big(\underline{\V{x}}_k^{(j)}\!, 1\big)
  \label{eq:beliefLegacy}
\end{equation}
and $\tilde{f}\big(\underline{\V{x}}_k^{(j)}\!, 0\big)\rmv=\rmv \underline{f}_k^{(j)} f_{\text{D}}\big(\underline{\V{x}}^{(j)}_{k}\big)$ with\vspace{0mm} $\underline{f}_k^{(j)} \rmv=\rmv \alpha_{\mathrm{n},k}^{(j)}\gamma_k^{(j)} /\underline{C}_k^{(j)}$. The constant $\underline{C}_k^{(j)}$ is given by $\underline{C}_k^{(j)}\rmv\triangleq\rmv \int \alpha_k^{(j)}\big(\underline{\V{x}}_k^{(j)}\!, 1\big)$ $\gamma_k^{(j)}\big(\underline{\V{x}}_k^{(j)}\!, 1\big) \mathrm{d}\underline{\V{x}}_k^{(j)} + \alpha_{\mathrm{n},k}^{(j)}\gamma_k^{(j)}\rmv$.



\section{Particle Flow Implementation of Messages}
\label{sec:PFMessages}
In nonlinear \ac{mot} scenarios, calculation of $\beta_k^{(j)}\big(a_k^{(j)}\big)$ in \eqref{eq:messageBeta} and $\tilde{f}\big(\underline{\V{x}}_k^{(j)}\!, \underline{r}_k^{(j)}\big)$ in \eqref{eq:beliefLegacy} related to legacy \ac{po} states as well as their counterparts related to new \ac{po} states cannot be performed in closed form. We propose a particle-based implementation, where a proposal \ac{pdf} is established by means of invertible particle flow (c.f. \ac{edh} method in \ref{sec:numericalImplementation}). This makes it possible to implement \eqref{eq:messageBeta} and \eqref{eq:beliefLegacy} (as well as their counterparts related to new \ac{po} states) by means of Monte Carlo integration and importance sampling, respectively. Contrary to an implementation where predicted object beliefs are used as proposal \ac{pdf} \cite{MeyBraWilHla:J17}, the number of particles and the computational complexity are strongly reduced. 




\subsection{Particle Flow and Measurement Evaluation}
\label{sec:samplingFlowEvaluation}

It is assumed that for the messages $\alpha_k^{(j)}\big(\underline{\V{x}}_k^{(j)},\underline{r}_k^{(j)}\big)$, $j\!\in\! \{1,\dots,j_{k-1}\}$ resulting from the prediction step, a Gaussian approximation with mean  $\underline{\V{x}}_{k}^{(j)\ast}$ and covariance matrix $\underline{\M{P}}_k^{(j)}\rmv\rmv$, i.e., $\alpha_k^{(j)}\big(\underline{\V{x}}_k^{(j)},1\big) \rmv \approx \tilde{\alpha}_{\mathrm{e},k}^{(j)} \hspace{1mm}  \mathcal{N}(\underline{\V{x}}_{k}^{(j)}\rmv\rmv;\underline{\V{x}}_{k}^{(j)\ast}\rmv\rmv,\underline{\M{P}}_k^{(j)})$, as well as a particle representation $\big\{\big(\underline{\V{x}}^{(j,i)}_{0,k},\underline{w}^{(j,i)}_{0,k}\big)\big\}_{i=1}^{N_\mathrm{s}}$ with $\sum^{N_{\mathrm{s}}}_{i = 1} \underline{w}^{(j,i)}_{0,k} =  \tilde{\alpha}_{\mathrm{e},k}^{(j)}$ are available\footnote{If only a particle representation $\big\{\rmv\big(\underline{\V{x}}^{(j,i)}_{0,k},\underline{w}^{(j,i)}_{0,k}\big)\rmv\big\}_{i=1}^{N_\mathrm{s}}$ of $\alpha_k^{(j)}\big(\underline{\V{x}}_k^{(j)},\underline{r}_k^{(j)}\big)$ is available, we can \vspace{-.15mm} calculate the mean $\underline{\V{x}}_{k}^{(j)\ast}$ and the covariance $\underline{\M{P}}_k^{(j)}$ of the Gaussian approximation as $\underline{\V{x}}_k^{(j)\ast} = \big(1/\tilde{\alpha}_{\mathrm{e},k}^{(j)}\big)  \sum_{i=1}^{N_\mathrm{s}}\underline{w}_{0,k}^{(j,i)} \underline{\V{x}}_{0,k}^{(j,i)} $ and $\underline{\M{P}}_k^{(j)} = \big(1/\tilde{\alpha}_{\mathrm{e},k}^{(j)}\big)  \sum_{i=1}^{N_\mathrm{s}} \underline{w}_{0,k}^{(j,i)}\big(\underline{\V{x}}_{0,k}^{(j,i)}-\underline{\V{x}}_k^{(j)\ast}\big)\big(\underline{\V{x}}_{0,k}^{(j,i)}-\underline{\V{x}}_k^{(j) \ast}\big)^{\rmv\T}\rmv\rmv\rmv$. If only a Gaussian approximation $\alpha_k^{(j)}\big(\underline{\V{x}}_k^{(j)},1\big) \rmv \approx \tilde{\alpha}_{\mathrm{e},k}^{(j)} \hspace{1mm}  \mathcal{N}(\underline{\V{x}}_{k}^{(j)}\rmv\rmv;\underline{\V{x}}_{k}^{(j)\ast}\rmv\rmv,\underline{\M{P}}_k^{(j)})$ is \vspace{-.15mm}available, we can obtain a particle representation $\big\{\big(\underline{\V{x}}^{(j,i)}_{0,k},\underline{w}^{(j,i)}_{0,k}\big)\big\}_{i=1}^{N_\mathrm{s}}$ by drawing particles $\underline{\V{x}}^{(j,i)}_{0,k}$, $i \rmv\in\rmv \{1,\dots,N_{\mathrm{s}} \}$ from $\mathcal{N}(\underline{\V{x}}_{k}^{(j)}\rmv\rmv;\underline{\V{x}}_{k}^{(j)\ast}\rmv\rmv,\underline{\M{P}}_k^{(j)})$ and setting the weights to $\underline{w}^{(j,i)}_{0,k} \rmv=\rmv \tilde{\alpha}_{\mathrm{e},k}^{(j)} /N_{\mathrm{s}}$.}. First, we compute an extended set $\big\{\big(\underline{\V{x}}^{(j,i)}_{1,k},\underline{w}^{(j,i)}_{1,k}\big)\big\}_{i=1}^{(m_k+1)N_{\mathrm{s}}}\rmv\rmv\rmv\rmv$, that consists of $N_{\mathrm{s}}$
particles and weights for each value \vspace{.2mm} of $a_k^{(j)} \in \{0,\dots,m_k\}$ and $j\!\in\! \{1,\dots,j_{k-1}\}$ as follows. For $a_k^{(j)}\! = \! 0$, we perform a zero flow by setting $\big\{\big(\underline{\V{x}}^{(j,i)}_{1,k},\underline{w}^{(j,i)}_{1,k}\big)\big\}_{i=1}^{N_{\mathrm{s}}}\! =\! \big\{\big(\underline{\V{x}}^{(j,i)}_{0,k},\underline{w}^{(j,i)}_{0,k}\big)\big\}_{i=1}^{N_{\mathrm{s}}}$. For $a_k^{(j)}\! =\! m\!\in\! \{1,\dots,m_k\}$, we perform an \ac{edh} particle flow as discussed in Section \ref{sec:numericalImplementation} by migrating particles $\big\{\underline{\V{x}}^{(j,i)}_{0,k}\big\}_{i=1}^{N_{\mathrm{s}}}$ to $\big\{\underline{\V{x}}^{(j,i)}_{1,k}\big\}_{i=m N_\mathrm{s}+1}^{(m+1)N_\mathrm{s}}$ based on the likelihood function $ f\big( \V{z}_{k}^{(m)} \rmv\big|\ist \underline{\V{x}}_{k}^{(j)} \big)$. 
(Note that measurement gating \cite{BarWilTia:B11} can be employed to reduce the number of measurements used for particle flow.)
By means of the invertible particle flow principle (c.f. \eqref{eq:invertibleMapping}) \cite{LiCoates:17}, the weights $\underline{w}_{1,k}^{(j,i)}$ corresponding to the migrated particles $\underline{\V{x}}^{(j,i)}_{1,k}$, $i \in \big\{m N_\mathrm{s}+1, \dots, (m+1)N_\mathrm{s}\big\}$ are obtained \vspace{1.5mm} as
\begin{equation}
  \underline{w}_{1,k}^{(j,i)} \rmv= \rmv\frac{\mathcal{N}(\underline{\V{x}}_{1,k}^{(j,i)};\underline{\V{x}}_{k}^{(j)\ast}\rmv\rmv\rmv,\underline{\M{P}}_k^{(j)}) \underline{\theta}^{(j)}_m\!}{\mathcal{N}(\underline{\V{x}}_{0,k}^{(j,i')};\underline{\V{x}}_{k}^{(j)\ast}\rmv\rmv\rmv,\underline{\M{P}}_k^{(j)})}\underline{w}_{0,k}^{(j,i')}\!, \hspace{2mm} m N_\mathrm{s}+1 \rmv\leq\rmv i \rmv\leq\rmv (m+1)N_\mathrm{s} \nn
  \vspace{.5mm}
\end{equation}
where $i' = i\rrmv\rmv\mod\rmv N_\mathrm{s}$ and $\underline{\theta}^{(j)}_m$ is the mapping factor (c.f. \eqref{eq:mappingFactor}). Note that the sets of weighted particles $\big\{\underline{\V{x}}_{1,k}^{(j,i)},\underline{w}_{1,k}^{(j,i)}\big\}_{i=m N_\mathrm{s}+1}^{(m+1)N_\mathrm{s}}$, $m \in \{1,\dots,m_k\}$ are based on different proposal \acp{pdf} but all represent $\alpha_{k}^{(j)}\rmv\big(\underline{\V{x}}_k^{(j)}\rmv\rmv,1\big)$.
As a result of migrating the particles $\underline{\V{x}}_{1,k}^{(j,i)}$ along the flow defined by the measurements $\V{z}_{k}^{(m)}$, $m \in \{1,\dots,m_k\}$, they are now at locations where the likelihood functions $f\big( \V{z}_{k}^{(m)} \rmv\big|\ist \underline{\V{x}}_{k}^{(j)} \big)$, $m N_\mathrm{s}+1 \rmv\leq\rmv i \rmv\leq\rmv (m+1)N_\mathrm{s}$ are significant. This makes it possible to approximate message passing operations accurately with a significantly reduced number of particles compared to a conventional particle-based implementation \cite{MeyBraWilHla:J17}.

Finally, based on weighted particles $\big\{\big(\underline{\V{x}}_{1,k}^{(j,i)},\underline{w}_{1,k}^{(j,i)}\big)\big\}_{i=a N_\mathrm{s}+1}^{(a+1)N_\mathrm{s}}$, the measurement evaluation step can now be performed by calculating an approximation $\tilde{\beta}_k^{(j)}(a)$ of the messages $\beta_k^{(j)}(a)$ in \eqref{eq:messageBeta} for all $j \!\in\! \{1,\dots,j_{k-1}\}$, $a \!\in\! \{0,\dots,m_k\}$ as
\begin{equation}
  \tilde{\beta}_k^{(j)}\big(a_k^{(j)} = a\big) \rmv=\rmv \rmv\sum_{i=a \rmv N_\mathrm{s}+1}^{(a+1)N_\mathrm{s}}\rrmv\rrmv\rmv q\big( \underline{\V{x}}^{(j,i)}_{1,k}\!, 1, a\rmv; \V{z}_{k} \big)\underline{w}_{1,k}^{(j,i)} + 1(a) \tilde{\alpha}_{\mathrm{n},k}^{(j)} \nn
\end{equation}
where  $\tilde{\alpha}_{\mathrm{n},k}^{(j)} \rmv=\rmv 1-\tilde{\alpha}_{\mathrm{e},k}^{(j)}$. For each new \ac{po} $m\!\in\! \{1,\dots,m_{k}\}$, sampling is performed by drawing particles $\overline{\V{x}}^{(m,i)}_{0,k}$, $i \rmv\in\rmv \{1,\dots,N_{\mathrm{s}} \}$ from $f_{\mathrm{b}}\big(\overline{\V{x}}^{(m)}_{k}\big)$ and setting the corresponding weights to $w^{(m,i)}_{0,k} \rmv=\rmv 1/N_{\mathrm{s}}$. Next, new particles $\big\{\overline{\V{x}}^{(m,i)}_{1,k}\big\}_{i=1}^{N_\mathrm{s}}$ are obtained from $\big\{\overline{\V{x}}^{(m,i)}_{0,k}\big\}_{i=1}^{N_{\mathrm{s}}}$ by means of particle flow based on $f\big(\V{z}_k^{(m)}  \big| \overline{\V{x}}^{(m)}_{k} \big)$. For these resulting particles $\big\{\overline{\V{x}}^{(m,i)}_{1,k}\big\}_{i=1}^{N_\mathrm{s}}$, weights are obtained, and then approximate messages $\tilde{\xi}_k^{(m)}\big(b_k^{(m)}\big)$ are calculated by performing the same steps as described above for the calculation of $\tilde{\beta}_k^{(j)}\big(a_k^{(j)}\big)$. These messages are used as an input for the iterative \ac{spa} for data association performed next (see \cite{WilLau:J14,MeyBraWilHla:J17,MeyKroWilLauHlaBraWin:J18} for details)\vspace{-1mm}.


\subsection{Measurement Update and Belief Calculation}
After the iterative loopy \ac{spa} for data association has converged, we have the resulting messages $\tilde{\kappa}_k^{(j)}\big(a_k^{(j)}\big)$, $j \!\in\! \{1,\dots,j_{k-1}\}$ and $\tilde{\iota}_k^{(m)}\big(b_k^{(m)}\big)$, $m \!\in\! \{1,\dots,m_k\}$ available. These messages are used to obtain an approximation $\tilde{\gamma}_k^{(j)}\big(\underline{\V{x}}_k^{(j)}\!, 1\big) $ of the messages $\gamma_k^{(j)}\big(\underline{\V{x}}_k^{(j)}\!, 1\big) $, $j \!\in\! \{1,\dots,j_{k-1}\}$ in \eqref{eq:messageGamma} as well as an approximation $\tilde{\varsigma}_k^{(m)}\big(\overline{\V{x}}_{k}^{(m)}\!, 1\big)$ of the messages $\varsigma_k^{(m)}\big(\overline{\V{x}}_{k}^{(m)}\!, 1\big)$, $m \!\in\! \{1,\dots,m_k\}$  in \cite[Section~IX]{MeyKroWilLauHlaBraWin:J18}.


Next, beliefs approximating the posterior \ac{pdf} of \acp{po} are computed by means of importance sampling. In particular, based on \eqref{eq:beliefLegacy}, we update the particle weights of the legacy \acp{po} $j \!\in\! \{1,\dots,j_{k-1}\}$\vspace{1mm} as 
\begin{equation}
  \underline{w}_{k}^{\mathrm{A}(j,i)} = \tilde{\gamma}_k^{(j)}\!\big(\underline{\V{x}}_{1,k}^{(j,i)}, 1\big) \ist \underline{w}_{1,k}^{(j,i)}, \iist\iist\iist 1 \rmv\leq\rmv i \rmv\leq\rmv (m_k+1)N_\mathrm{s} \nn
  \vspace{.5mm}
\end{equation}
and set $\underline{w}_{k}^{\mathrm{B}(j)} = \tilde{\alpha}_{\mathrm{n},k}^{(j)} \tilde{\gamma}_k^{(j)}$. 
Recall that for each $a\!\in\! \{0,1,\dots,m_k\}$, the set $\big\{\big(\underline{\V{x}}_{1,k}^{(j,i)},\underline{w}_{k}^{\mathrm{A}(j,i)}\big)\big\}_{i=aN_\mathrm{s}+1}^{(a+1)N_\mathrm{s}}$ is a particle representation of $\alpha_k^{(j)}\big(\underline{\V{x}}_k^{(j)}\!, 1\big) \ist \gamma_k^{(j)}\big(\underline{\V{x}}_k^{(j)}\!, 1\big)$ that is based on a different proposal \ac{pdf} as discussed in Section~\ref{sec:samplingFlowEvaluation}.

Considering the fact that an existent object can generate at most one measurement, we choose the particle representation that results in the largest sum of weights, i.e., $\big\{\big(\underline{\V{x}}_{1,k}^{(j,i)},\underline{w}_{k}^{\mathrm{A}(j,i)}\big)\big\}_{i=a' N_\mathrm{s}+1}^{(a'+1)N_\mathrm{s}}$ with 
\begin{equation}
a'= \argmax_{a\in\{0,\dots,m_k\}} \rmv\rmv \Bigg\{ \rmv \sum_{i=aN_\mathrm{s}+1}^{(a+1)N_\mathrm{s}} \hspace{-2mm}\underline{w}_{k}^{\mathrm{A}(j,i)} \Bigg\}. \nn
\vspace{1mm}
\end{equation}
In addition, we use the weighted particles $\big\{\big(\underline{\V{x}}_{1,k}^{(j,i)},\underline{w}_{k}^{\mathrm{A}(j,i)}\big)\big\}_{i=1}^{N_\mathrm{s}}$ to ensure that the combined proposal \ac{pdf}\footnote{Using these two particle sets corresponds to a combined proposal \ac{pdf} that consists of two equally-weighted components. The first component is based on particle flow defined by $a'$ and the second component is $\alpha_k^{(j)}\big(\underline{\V{x}}_k^{(j)},1\big)$. } is less informative than the final belief $\tilde{f}\big(\underline{\V{x}}_k^{(j)}\!, 1\big)$. More specifically, a new set of $2N_{\mathrm{s}}$ particles and \vspace{.1mm} weights $\big\{\big(\underline{\V{x}}_{1,k}^{(j,i)},\underline{w}_{k}^{\mathrm{A}(j,i)}\big)\big\}_{i=1}^{2N_\mathrm{s}}$ is obtained by taking the union \vspace{-.3mm} of the sets $\big\{\big(\underline{\V{x}}_{1,k}^{(j,i)},\underline{w}_{k}^{\mathrm{A}(j,i)}/2\big)\big\}_{i=1}^{N_\mathrm{s}}$ and $\big\{\big(\underline{\V{x}}_{1,k}^{(j,i)},\underline{w}_{k}^{\mathrm{A}(j,i)}/2\big)\big\}_{i=a'N_\mathrm{s}+1}^{(a'+1)N_\mathrm{s}}$. For this new set of \vspace{.4mm} particles, normalized weights are calculated as 
\begin{equation}
  \underline{w}_{k}^{(j,i)} = \frac{\underline{w}_{k}^{\mathrm{A}(j,i)}}{\sum_{i'=1}^{2N_\mathrm{s}}\underline{w}_{k}^{\mathrm{A}(j,i')}+\underline{w}_{k}^{\mathrm{B}(j)}}, \quad 1 \rmv\leq\rmv i \rmv\leq\rmv 2N_\mathrm{s}. 
  \label{eq:weightNormLegacy}
\end{equation}
Note that denominator of \eqref{eq:weightNormLegacy} corresponds to $\underline{C}_k^{(j)}$ in \eqref{eq:beliefLegacy}. The resulting particles and weights $\big\{\big(\underline{\V{x}}_{1,k}^{(j,i)},\underline{w}_{k}^{(j,i)}\big)\big\}_{i=1}^{2N_\mathrm{s}}$ represent the belief $\tilde{f}\big(\underline{\V{x}}_k^{(j)}\!, 1 \big)$ of legacy \ac{po} $j \rmv\in\rmv \{1,\dots,j_{k-1}\}$. 

From these particles and weights $\big\{\big(\underline{\V{x}}_{1,k}^{(j,i)},\underline{w}_{k}^{(j,i)}\big)\big\}_{i=1}^{2N_\mathrm{s}}$, we can calculate an approximation of the legacy \ac{po}'s existence probability as $\underline{p}_{k}^{\mathrm{e}(j)} = \sum_{i=1}^{2N_\mathrm{s}}\underline{w}_{k}^{(j,i)}$ as well as an approximation of its \ac{mmse} state estimate $\hat{\underline{\V{x}}}_k^{(j)}$ in \eqref{eq:mmseEst} as
\begin{equation}
  \hat{\underline{\V{x}}}_k^{(j)} = \frac{1}{\underline{p}_{k}^{\mathrm{e}(j)}}\sum_{i=1}^{2N_\mathrm{s}}\underline{w}_{k}^{(j,i)}\underline{\V{x}}_{1,k}^{(j,i)}. \nn
  \vspace{.5mm}
\end{equation}
Finally, to reduce the number of particles to $N_{\mathrm{s}}$ and avoid particle degeneracy, a particle resampling step is performed \cite{AruMasGorCla:02}. For new \acp{po} $m \rmv\in\rmv \{1,\dots,m_k\}$, particles representing beliefs as well as estimates of the existence probabilities and states are calculated by performing similar steps as discussed above for legacy \acp{po}.

\section{Simulation Results}
\label{sec:simResults}

We consider a 3-D tracking scenario with eight objects and $200$ time steps. The object states at time $k$ consist of 3-D position and velocity, i.e., $\RV{x}^{(j)}_{k} = [\rv{x}^{(j)}_{1,k}\iist\rv{x}^{(j)}_{2,k}\iist\rv{x}^{(j)}_{3,k}\iist\rv{\dot{x}^{(j)}_{1,k}}\iist\rv{\dot{x}}^{(j)}_{2,k}\iist\rv{\dot{x}}^{(j)}_{3,k}]^{\text{T}}$ and evolve according to a constant-velocity model \cite[Sec.~6.3.2]{BarRonKir:01} with driving noise variance $0.01\ist$m$^2$/s$^4$.  The region of interest (ROI) is $[-500\ist\text{m}, \ist 500\ist\text{m} ]  \times [-500\ist\text{m}, \ist 500\ist\text{m}] \times [-500\ist\text{m}, \ist 0\ist\text{m}]$. Objects appear at $k \rmv\in\rmv \{1,10,20,30,40,50,60,70\}$ and disappear at $k \rmv \in \{130,$ $140, \rmv150,\rmv160,\rmv170,\rmv180,\rmv190\}$; their tracks intersect at the ROI center. 

\ac{tdoa} measurements are generated by two arrays that consist of five receivers and have the same geometry \cite{TesMeyBee:20}; the arrays are located at $[250 \ist\ist\ist 0 \ist\ist -\rmv10]^{\T}$ and $[0 \ist\ist\ist 250 \ist\ist -\rmv10]^{\T}\rmv\rmv$. The same six pairs of receivers are selected on each array to generate noisy \ac{tdoa} measurements. This means that each joint measurement $\RV{z}_{k}^{(m)}$ obtained by the two arrays at time $k$ consists of $12$ individual noisy \acp{tdoa} $\rv{z}^{(m)}_{l,k}$, $l \rmv\in\rmv \{1,\dots,12\}$. In particular, the noisy \ac{tdoa} $\rv{z}^{(m)}_{l,k}$ of an object with state $\RV{x}^{(j)}_{k}$ acquired by a pair of receivers $(s_l,t_l)$ at positions $\V{p}_{s_l}$ and $\V{p}_{t_l}\rmv$, is modeled \vspace{1.3mm} as
\begin{align}
\rv{z}^{(m)}_{l,k} &= \frac{1}{c} \Big( \big\| \big[\rv{x}^{(j)}_{1,k}\ist\ist\ist \rv{x}^{(j)}_{2,k}\ist\ist\ist \rv{x}^{(j)}_{3,k}\big]^{\text{T}} - \V{p}_{s_l}\big\| \nn\\[.8mm]
&\hspace{11mm}- \big\| \big[\rv{x}^{(j)}_{1,k}\ist\ist\ist \rv{x}^{(j)}_{2,k}\ist\ist\ist \rv{x}^{(j)}_{3,k}\big]^{\text{T}} - \V{p}_{t_l} \big\|\Big)+\rv{v}_{l,k}^{(m)} \nn\\[-3.5mm]
\nn
\end{align}
where $c=1500\ist$m/s is the propagation speed, and $\rv{v}_{l,k}^{(m)}$ is additive zero-mean Gaussian noise with standard deviation $\sigma_{\rv{v}} = 3\times 10^{-6}$s that is assumed statistically independent across $l$, $k$, and $m$. The probability of detection is $p_\text{d}=0.9$ and the mean number of clutter measurements is $\mu_\text{c}=1$. The clutter is assumed statistically independent across $l$, i.e., $f_{\text{c}}\big( \V{z}^{(m)}_{k} \big) \rmv=\rmv \prod^{12}_{l = 1} f_{\text{c}}\big( z^{(m)}_{l,k} \big) $. The individual clutter \acp{pdf} $f_{\text{c}}\big( z^{(m)}_{l,k} \big)$ are assumed uniform on $\frac{1}{c} \big[-\| \V{p}_{s_l} - \V{p}_{t_l}\|,\| \V{p}_{s_l} - \V{p}_{t_l}\|\big] $.
The \ac{pdf} for object birth $f_\text{b}(\cdot)$ is uniform on the ROI, and the mean number of newborn objects is $\mu_\text{b} = 0.011$. The survival probability is $p_\mathrm{s}=0.999$. The object declaration threshold is set to $P_{\text{th}} \rmv=\rmv 0.5$ and the pruning threshold to $P_{\text{pr}} \rmv=\rmv 10^{-4}$.

We compare the proposed \acp{spa}-based \ac{mot} algorithm with embedded particle flow (``SPA-PF'') with two reference methods also based on the  \ac{spa} based multiobject tracking framework in \cite{MeyKroWilLauHlaBraWin:J18}. The first method (``SPA-PM'') uses predicted object beliefs as proposal \ac{pdf} \cite{MeyBraWilHla:J17}. The second (yet unpublished) method (``SPA-UT'') uses the unscented transformation to calculate an informative proposal \ac{pdf} \cite{MerDouFre:00}. We set $N_{\mathrm{s}} \rmv\in\rmv \{500000, 10000\}$ for SPA-PM, $N_{\mathrm{s}} \rmv\in\rmv \{10000, 500 \}$ for SPA-UT, and $N_{\mathrm{s}} \rmv\in\rmv \{1000, 100 \}$ for SPA-PF and performed 1000 simulation runs. For all three methods, we calculate messages and beliefs related to newborn objects by increasing this number of particles by a factor of 20.

\begin{figure}[t!]
\centering

\psfrag{s01}[b][b]{\color[rgb]{0.15,0.15,0.15}\setlength{\tabcolsep}{0pt}\begin{tabular}{c}\raisebox{1.5mm}{MOSPA}\end{tabular}}%
\psfrag{s04}[t][t]{\color[rgb]{0.15,0.15,0.15}\setlength{\tabcolsep}{0pt}\begin{tabular}{c}\raisebox{-1.5mm}{time step $k$}\end{tabular}}%
%
\color[rgb]{0.15,0.15,0.15}%
%
\psfrag{x01}[t][t]{$0$}%
\psfrag{x02}[t][t]{$50$}%
\psfrag{x03}[t][t]{$100$}%
\psfrag{x04}[t][t]{$150$}%
\psfrag{x05}[t][t]{$200$}%
%
\psfrag{v01}[r][r]{$0$}%
\psfrag{v02}[r][r]{$10$}%
\psfrag{v03}[r][r]{$20$}%
\psfrag{v04}[r][r]{$30$}%
\psfrag{v05}[r][r]{$40$}%
\psfrag{v06}[r][r]{$50$}%
\psfrag{v07}[r][r]{$60$}%

\psfrag{BPF-500000}[l][l][.84]{SPA-PM-500000}%
\psfrag{BPF-10000}[l][l][.84]{SPA-PM-10000}%
\psfrag{UPF-10000}[l][l][.84]{SPA-UT-10000}%
\psfrag{UPF-500}[l][l][.84]{SPA-UT-500}%
\psfrag{PFPF-1000 (proposed)}[l][l][.84]{SPA-PF-1000 (proposed)}%
\psfrag{PFPF-100 (proposed)}[l][l][.84]{SPA-PF-100 (proposed)}%

\includegraphics[height=50mm, width=72mm]{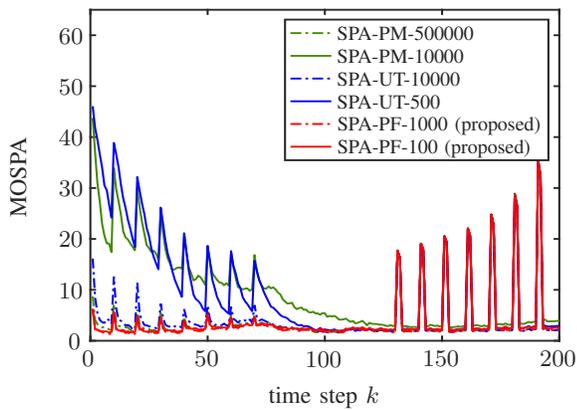}
\vspace{2mm}
\caption{Multiobject tracking performance characterized by the MOSPA error (cutoff parameter $C = 50$) versus time $k$. Note that at the time step where the objects disappear, i.e., $k = 130, 140, \dots, 190$, the performance of all the simulated methods coincides. }
\vspace{0mm}
\label{fig:trackingPerformance}
\end{figure} 

We measure the performance of the various algorithms by the Euclidean distance based \ac{ospa} metric with cutoff parameter $C=50$ \cite{SchVoVo:J08}. Fig. \ref{fig:trackingPerformance} shows the \ac{mospa} error---averaged over 1000 simulation runs---of all methods versus time $k$.  The average computation time per time step $k$ for a MATLAB implementation on a single core of an Intel Xeon Gold 5222 CPU was measured as 10.19s for SPA-PM-500000; 0.08s for SPA-PM-10000; 1.62s for SPA-UT-10000; 0.04s for SPA-UT-500; 0.55s for SPA-PF-1000; and 0.12s \vspace{0mm} for SPA-PF-100.

All simulated methods yield error peaks at time steps where objects appear and disappear. However, it can be seen that the proposed method outperforms the two reference methods at almost all time steps (while the performance for all the methods at the object disappearances coincides). In particular, the proposed SPA-PF yields a significantly reduced \ac{mospa} error at time steps after object appearance. Increasing the number of particles does improve the performance for SPA-PM and SPA-UT, but this comes at the cost of increased computational complexity and memory requirements.
Notably, the performance of SPA-PF is not affected by an increase of the number of particles from $N_{\mathrm{s}} = 100$ to $N_{\mathrm{s}} = 1000$. It can be concluded that the proposed SPA-PF can outperform state-of-the-art referenced methods that have higher computational complexity and memory requirements due to a larger number of particles.

\acresetall 

\section{Conclusion}
\label{sec:conclusion}

We presented a graph-based Bayesian method for challenging nonlinear and high-dimensional \ac{mot} problems that relies on particle filtering. Particle degeneracy is avoided by performing operations on the graph using particle flow. Our numerical results demonstrate reduced computational complexity and memory requirements as well as favorable detection and estimation accuracy in a 3-D \ac{mot} scenario. The introduced approach is expected to be particularly appealing for large scale machine perception problems \cite{CadCarCar:16}. Possible directions of future research also include graph-based methods with embedded particle flow for simultaneous localization and object tracking \cite{MeyHliHla:J16}, extended object tracking \cite{MeyWin:J20}, and information-seeking control \cite{MeyWymFroHla:J15}.

\renewcommand{\baselinestretch}{1.04}
\selectfont
\bibliographystyle{IEEEtran}
\bibliography{StringDefinitions,IEEEabrv,Papers,Books,Temp}









\end{document}